\documentstyle[12pt]{article}
\input epsf
\font\upright=cmu10 scaled\magstep1
\font\sans=cmss10

\newcommand{\F}{{\cal F}}
\newcommand{\ssf}{\sans}
\newcommand{\Z}{\hbox{\upright\rlap{\ssf Z}\kern 2.7pt {\ssf Z}}}
\newcommand{\R}{\hbox{\upright\rlap{I}\kern 1.7pt R}}
\newcommand{\C}{\hbox{\upright\rlap{I}\kern -1.9pt C}}
\newcommand{\la}{\label}

\newcommand{\be}{\begin{equation}}
\newcommand{\ee}{\end{equation}}
\newcommand{\ba}{\begin{eqnarray}}
\newcommand{\ea}{\end{eqnarray}}
\newcommand{\bastar}{\begin{eqnarray*}}
\newcommand{\eastar}{\end{eqnarray*}}
\begin{document}

\begin{titlepage}
 
\begin{center}
\vskip 4.0cm
{\bf \large CLASS OF EXACT SOLUTIONS OF \\ \vskip 0.3cm
            THE SU(3) SKYRME MODEL \\}
\end{center}
 
\vskip 2.0cm
 
\begin{center}
{\bf Wang-Chang Su$^{*}$ }  \\
\vskip 0.3cm
{\it Department of Physics \\
     National Chung-Cheng University \\
     Chia-Yi 621, Taiwan } \\
\end{center}
 
\vskip 0.5cm

{\rm
Recently, Hirayama and Yamashita have presented an ansatz that allows us to construct a class of solutions for the $SU(2)$ Skyrme model.
Though these solutions are not solitonic, they provide us with an example on how the plane wave solutions arise in nonlinear field theories.
In this paper, we investigate the applicability of the ansatz for the $SU(3)$ Skyrme model.
We explicitly construct a class of solutions for the $SU(3)$ model, which in the simplest circumstance is reduced to a combination of the plane waves and Weierstrass elliptic functions. 
We also discuss some properties of these solutions. 
For example, the intrinsic structure of these solutions is found to describe an asymmetrical top rotating in the complex three-dimensional space.  
}

\noindent\vfill
 
\begin{flushleft}
\rule{5.1 in}{.007 in} \\
$^{*}$ \hskip 0.2cm {\small  E-mail: \bf suw@phy.ccu.edu.tw } \\
\end{flushleft}

\end{titlepage}


\section{\la{INTRODUCTION}
             INTRODUCTION}

Quantum Chromodynamics (QCD) is known to be the theory of strong interactions.
At the high-energy, the theory describes weakly interacting quarks and gluons, and can be solved perturbatively thanks to asymptotic freedom.
At the low-energy, the QCD theory is supposed to describe the interactions of mesons, baryons and probably glueballs, but it becomes strongly coupled.
The theory is difficult to solve from the first principles at the low-energy limit.
Nevertheless, the chiral symmetry properties of QCD in the zero quark mass limits do lead us to effective lagrangian for the description of strong interactions at the low-energy.
Among them, the Skyrme model \cite{Skyrme} is the most successful one, which contains terms compatible with the symmetries of QCD and at most quadratic in time derivative \cite{tHooft,Witten1}. 
Based on the Skyrme model, considerable progress has been made in understanding the low-energy properties of QCD, {\it e.g.} the dynamics of mesons and baryons.
It is shown that the predictions of the Skyrme model agree well with the experimental data \cite{Witten2,Jackson}. 
Furthermore, the recent numerical investigation of the Skyrme model has revealed the intricate and fascinating structures of the model, such as the fullerene structures of the higher baryon number solitons \cite{Battye}.

We first briefly review some aspects of the $SU(N)$ ($N \ge 2$) Skyrme model. 
The Skyrme model is a nonlinear theory of the Skyrme field $U(x)$.
The nonlinear constraints on the field $U(x)$ are unitarity and unimodularity conditions, 
that is, the Skyrme field $U(x)$ is a matrix in $SU(N)$ group. 
The invariant lagrangian density, compatible with the chiral symmetries of QCD, takes the form
\ba
{\cal L} = 
\frac{F^2_\pi}{16} \, \mbox{Tr}
\left( \partial_\mu U^\dagger \, \partial_\mu U \right) +
\frac{1}{32 e^2} \, \mbox{Tr}
\left( \left[ U^\dagger \, \partial_\mu U, U^\dagger \, \partial_\nu U \right]^2 \right).
\la{skyrme1}
\ea
Here we omit the Wess-Zumino-Witten term since it leads to no equation of motion, thus is irrelevant to our discussion.
The chiral symmetry breaking terms are not considered, either.
In the lagrangian density (\ref{skyrme1}), $e$ is a dimensionless parameter and $F_\pi$ is known as the pion decay constant.
The first term in (\ref{skyrme1}) is the standard nonlinear chiral lagrangian, where as the second term is called the Skyrme term.
The first term alone is insufficient to demonstrate the existence of classical stable topological solutions, as suggested by Derrick's scaling theorem \cite{derrick}.
The Skyrme term is thus introduced to prevent the instability of field configurations.
As a result, the model (\ref{skyrme1}) containing  both nonlinear chiral and Skyrme terms support the stable topological solitons.

We then introduce the left-invariant Maurer-Cartan covariant vector $\mbox{\boldmath$L$}_\mu$, which is an $SU(N)$ Lie-algebra valued current, as
\ba
\mbox{\boldmath$L$}_\mu \equiv
L_\mu^\alpha \, T_\alpha = 
\frac{1}{i} \, U^\dagger \, \partial_\mu U \, ,
\la{leftcurrent}
\ea
where $T_\alpha$ for \( \alpha = 1, \dots , N^2-1 \) are the generators of $SU(N)$ in the fundamental representation. 
By construction, the covariant vector $\mbox{\boldmath$L$}_\mu (x)$ acts as a pure gauge connection.
It satisfies the zero curvature condition, namely, the Maurer-Cartan identity
\ba
\partial_{\mu} \mbox{\boldmath$L$}_\nu -\partial_{\nu} \mbox{\boldmath$L$}_\mu
= -i \left[ \mbox{\boldmath$L$}_\mu , \mbox{\boldmath$L$}_\nu \right].
\la{MCidentity}
\ea
In terms of the Maurer-Cartan covariant vector (\ref{leftcurrent}), the Skyrme lagrangian density in (\ref{skyrme1}) is rewritten as
\ba
{\cal L} = 
\frac{F^2_\pi}{16} \, \mbox{Tr}
\left( \mbox{\boldmath$L$}_\mu \, \mbox{\boldmath$L$}_\mu \right) +
\frac{1}{32 e^2} \, \mbox{Tr}
\left( \left[ \mbox{\boldmath$L$}_\mu , \mbox{\boldmath$L$}_\nu \right]^2 \right).
\la{skyrme2}
\ea

The Skyrme model (\ref{skyrme2}) or equivalently (\ref{skyrme1}) admits stable solitonic solutions. known as skyrmions, that can be characterized by an integer-valued topological number.
It is suggested that this topological number be identified with the baryon number of the nucleon \cite{Skyrme}.
Explicitly, the conservation of the baryon number is associated to the topological charge
\ba
\mbox{\boldmath$Q$} = \frac{i}{48 \pi^2} \,
\epsilon_{i j k} \int d^3 x \, \mbox{Tr}
\left( 
\mbox{\boldmath$L$}_i \,
\left[ \mbox{\boldmath$L$}_j, \mbox{\boldmath$L$}_k \right]
\right).
\la{baryonnumber}
\ea
In the above formula, the charge $\mbox{\boldmath$Q$}$ takes integer values on each topologically distinct class of field configurations, \( U : \R^3 \to S^3 \).
Mathematically, such mapping falls into nontrivial homotopy class, \( \pi_3 (S^3) = \Z \).
It is readily to check that the expression for $\mbox{\boldmath$Q$}$ in (\ref{baryonnumber}) is indeed a conserved quantity.
To this aim, we write the corresponding conserved topological current:
\ba
\mbox{\boldmath$J$}_\mu = \frac{i}{48 \pi^2} \,
\epsilon_{\mu \nu \sigma \rho} \, \mbox{Tr}
\left( 
\mbox{\boldmath$L$}_\nu
\left[ \mbox{\boldmath$L$}_\sigma, \mbox{\boldmath$L$}_\rho \right]
\right).
\la{baryoncurrent}
\ea
Then, the conservation of the topological charge $\mbox{\boldmath$Q$}$, in which the current component $\mbox{\boldmath$J$}_0$ plays the role of the charge density, is just a consequence of the conservation law of the topological current \( \partial_\mu \mbox{\boldmath$J$}_\mu = 0 \).

Actually, the Skyrme model has much richer soliton spectrum than one's expectation.
It was recently shown by Cho \cite{Cho} that, with an appropriate choice of the Skyrme field, the Skyrme model is a theory of self-interacting non-Abelian monopoles and admits knot-like solitons. 
These solitons are very similar to the topological knots in the Faddeev-Skyrme model \cite{Faddeev}, but are different from the well-known skyrmions.

Moreover, the Euler-Lagrange's equation of the Skyrme model (\ref{skyrme2}) satisfied by each topological field configuration can be easily derived from the standard variational principle. 
It reads
\ba
\partial_{\mu} 
\left(
F^2_\pi \, \mbox{\boldmath$L$}_\mu +
\frac{1}{e^2} \, \left[ \mbox{\boldmath$L$}_\nu , 
\left[ \mbox{\boldmath$L$}_\mu , \mbox{\boldmath$L$}_\nu \right] \right]
\right) = 0.
\la{Motion}
\ea

Because of the nonlinear characteristics of the Skyrme model, only a few analytic solutions to the Lagrange's equation of motion (\ref{Motion}) are known. 
Take the $SU(2)$ Skyrme model as an example, it was pointed out \cite{Skyrme} that the following choice of the Skyrme field in the covariant vector $\mbox{\boldmath$L$}_\mu (x)$ (\ref{leftcurrent})
\ba
U (x) = U (k \cdot x) 
\la{solution1}
\ea
is a solution of Eq. (\ref{Motion}) provided that the $k_\mu$ is a light-like four-momentum, that is, $k^2 = 0$.
In addition, the static Skyrme configuration with the spherically symmetric ansatz
\ba
U (\vec{r})= \cos\theta (r) + i \, \hat{r}_a \tau_a \, \sin\theta (r),
\la{solution2}
\ea
with $\hat{r} = \vec{r}/r$, renders a nontrivial solution of unit topological charge of the field equation (\ref{Motion}). 
Here, the profile function $\theta (r)$ is what minimizes the static energy functional of the model \cite{Witten2}.
As far as the analytic solutions of the $SU(N)$ $(N > 2)$ Skyrme model concerned, the harmonic map ansatz of $S^2$ to $CP^{N-1}$ that is the generalization of (\ref{solution2}) allows us to construct radially symmetric solutions with nontrivial topological charge \cite{Ioannidou}. 
In the general case, these solutions depend on $N-1$ profile functions that have to be determined numerically. 

Recently, a class of exact solutions for the $SU(2)$ Skyrme model is presented by Hirayama and Yamashita \cite{Hirayama}. 
In their result, the covariant field $\mbox{\boldmath$L$}_\mu (x)$ (\ref{leftcurrent}) is constructed explicitly, yet the final expression for the Skyrme field $U(x)$ is symbolical since it contains an ordering operation. 
In contrast to the previously found solutions, it is noted that these solutions depend on the both couplings $F_\pi$ and $e$ of the model (\ref{skyrme1}).
These solutions are found not to be solitonic, but with wave characteristics in the Minkowski space. 
At any rate, the solutions supply us with an example on how the superposition of wave solutions arises in nonlinear field theories. 

In this paper, we examine the applicability of the Hirayama-Yamashita ansatz for the $SU(3)$ Skyrme model \cite{Hirayama}, 
Based on the ansatz, we generalize their findings. 
The covariant field $\mbox{\boldmath$L$}_\mu (x)$ (\ref{leftcurrent}) is given explicitly.
A general class of solutions for the $SU(3)$ model is constructed, in which the simplest form of the solution is written as a complicated combination of the plane waves and Weierstrass elliptic functions. 
Because the derivation of the general solutions at some stages is quite involved, we shall report the final results only, but skip the detailed calculations.
Furthermore, we discuss some properties of these solutions. 
For example, we find that the intrinsic structure of the solutions represents an asymmetrical top rotating under the influence of the external torque in the complex three-dimensional space.    
In other words, they describe the rotational motion of an $SO(3,\C)$ top.
We also show that the $SU(3)$ solutions established here can be easily reduced to the $SU(2)$ solutions in \cite{Hirayama}.   
This paper is organized as follows. 
In Sec. \ref{HIRAYAMA-YAMASHITA ANSATZ}, we discuss the Hirayama-Yamashita ansatz. 
In Sec. \ref{SOLUTION OF THE SU(3) SKYRME MODEL}, we obtain a class of exact solutions for the $SU(3)$ Skyrme model using the ansatz. 
We discuss the extreme case of the solutions and conclude the paper in Sec. \ref{DISCUSSIONS AND CONCLUSIONS}.

 
\section{\la{HIRAYAMA-YAMASHITA ANSATZ}
             HIRAYAMA-YAMASHITA ANSATZ}

In the paper of Hirayama and Yamashita \cite{Hirayama}, a class of exact solutions for the $SU(2)$ Skyrme model is presented. 
The solutions provide us with an example on how the plane wave solutions can arise in nonlinear field theories. 
In this section, we review the ansatz for the general $SU(N)$ Skyrme model and obtain the simplified equations in which the covariant vector $\mbox{\boldmath$L$}_\mu (x)$ (\ref{leftcurrent}) satisfies.
In brief, we are seeking the solution of the Skyrme field (\ref{Motion}) of the following form:
\ba
U(x) = U(\xi, \eta),
\la{HYansatz}
\ea
where we define $\xi = k \cdot x$ and $\eta = l \cdot x$.
The two constant four-vectors $k_\mu$ and $l_\mu$ are light-like and arbitrary, that is, they satisfying \( k^2 = l^2 = 0 \) and \( k \cdot l \neq 0 \).
The Hirayama-Yamashita solution (\ref{HYansatz}) can be regarded as a generalization of (\ref{solution1}). 
Using the solution (\ref{HYansatz}), the left current $\mbox{\boldmath$L$}_\mu (x)$ defined in (\ref{leftcurrent}) is written as
\ba
\mbox{\boldmath$L$}_\mu (x) = 
k_{\mu} \, \mbox{\boldmath$A$}(\xi, \eta) +
l_{\mu} \, \mbox{\boldmath$\Lambda$}(\xi, \eta).
\la{leftcurrent1}
\ea
Here, $\mbox{\boldmath$A$} (\xi, \eta)$ and $\mbox{\boldmath$\Lambda$} (\xi, \eta)$ are two Lie-algebra valued fields. They are defined, respectively, by
\ba
\mbox{\boldmath$A$} (\xi, \eta)
&=& A_\alpha (\xi, \eta) \, T_\alpha 
\equiv \frac{1}{i} \, U^{\dagger}(\xi, \eta) \, {\partial_\xi} U(\xi, \eta), 
\la{Adefinition} \\
\mbox{\boldmath$\Lambda$} (\xi, \eta)
&=& \Lambda_\alpha (\xi, \eta) \, T_\alpha 
\equiv
\frac{1}{i} \, U^{\dagger}(\xi, \eta) \, {\partial_\eta} U(\xi, \eta),
\la{Lambdadefinition}
\ea
where \( \alpha = 1,2, \dots, N^2-1 \).
The shorthand notations \( {\partial_\xi} = \frac{\partial}{\partial \xi} \) and
\( {\partial_\eta} = \frac{\partial}{\partial \eta} \) are used.
In terms of the both Lie-algebra valued fields $\mbox{\boldmath$A$}(\xi, \eta)$ and $\mbox{\boldmath$\Lambda$}(\xi, \eta)$, the Maurer-Cartan identity (\ref{MCidentity}) and the Lagrange's field equation (\ref{Motion}) can be expressed as
\ba
{\partial_\eta} \mbox{\boldmath$A$} - {\partial_\xi} \mbox{\boldmath$\Lambda$} =
- i \left[ \mbox{\boldmath$\Lambda$}, \mbox{\boldmath$A$} \right],
\label{MCidentity1}
\ea
\ba
{\partial_\eta}
\left( 
\sigma \, \mbox{\boldmath$A$} -
\left[ \left[ \mbox{\boldmath$A$}, \mbox{\boldmath$\Lambda$} \right], \mbox{\boldmath$A$} \right]
\right) +
{\partial_\xi}
\left(
\sigma \, \mbox{\boldmath$\Lambda$} -
\left[ \left[ \mbox{\boldmath$\Lambda$}, \mbox{\boldmath$A$} \right], \mbox{\boldmath$\Lambda$} \right]
\right) = 0,
\label{Motion1}
\ea
where \( \sigma = \frac{F^2_\pi e^2}{(k \cdot l)} \) is a combined dimensionless parameter.

Now, we further simplify (\ref{MCidentity1}) and (\ref{Motion1}) by assuming that the field $\mbox{\boldmath$\Lambda$}$ is a constant Lie-algebra valued element.
Consequently, both equations are reduced to
\ba
{\partial_\eta} \mbox{\boldmath$A$} =
- i \left[ \mbox{\boldmath$\Lambda$}, \mbox{\boldmath$A$} \right], 
\label{MCidentityansatz} 
\ea
\ba
\left[
\sigma \, \mbox{\boldmath$A$} -
\left[ \mbox{\boldmath$A$}, \left[ \mbox{\boldmath$\Lambda$}, \mbox{\boldmath$A$} \right] \right] +
\left[ \mbox{\boldmath$\Lambda$}, i \, {\partial_\xi} \mbox{\boldmath$A$} \right],
\mbox{\boldmath$\Lambda$}
\right] = 0.
\label{Motionansatz}
\ea
In next section, we shall first use the simplified Maurer-Cartan identity (\ref{MCidentityansatz}) to obtain a general solution for the Lie-algebra valued field $\mbox{\boldmath$A$}(\xi,\eta)$, which is then substituted into the Lagrange's equation (\ref{Motionansatz}) for the determination of its final expression.


\section{\la{SOLUTION OF THE SU(3) SKYRME MODEL}
             SOLUTION OF THE SU(3) SKYRME MODEL}

In this section, a class of solutions of $SU(3)$ Skyrme model is established using the Hirayama-Yamashita ansatz as discussed in the previous section.
That is to say, we shall solve for the Lie-algebra valued field $\mbox{\boldmath$A$}(\xi, \eta)$ satisfying both (\ref{MCidentityansatz}) and (\ref{Motionansatz}) provided that the  $\mbox{\boldmath$\Lambda$}$ is a Lie-algebra valued constant.

The group $SU(3)$ has eight generators $T_\alpha$ \( (\alpha = 1~{\rm to}~8) \), obeying
the multiplication law 
\ba
T_\alpha \, T_\beta = \frac{1}{2}
\left( 
\frac{1}{3} \delta_{\alpha \beta} I_{(3)} + 
\left( d_{\alpha\beta\gamma} + i f_{\alpha\beta\gamma} \right) T_\gamma
\right) 
\ea
where $I_{(3)}$ is the three-dimensional unit matrix. 
The $f_{\alpha\beta\gamma}$ are totally antisymmetric and $d_{\alpha\beta\gamma}$ are totally symmetric under interchange of any two indices. 
The generators in the fundamental representation are given as \( T_\alpha = \frac{1}{2} \lambda_\alpha \), where $\lambda_\alpha$ are the standard Gell-Mann matrices. 

\subsection{The Eigenvalue Equation}

Since the simplified Maurer-Cartan equation (\ref{MCidentityansatz}) is linear in the Lie-algebra valued field $\mbox{\boldmath$A$} (\xi, \eta)$, we write the solution in the form
\ba 
\mbox{\boldmath$A$} (\xi,\eta) = 
e^{- \frac{i}{2} \omega \, \eta} \mbox{\boldmath$A$} (\xi) .
\la{Asolution1}
\ea
Then the substitution of this solution (\ref{Asolution1}) into the differential equation (\ref{MCidentityansatz}) results in a set of linear homogeneous equation for the hermitian matrix $\left( \Lambda_\gamma \F_\gamma \right)$ as
\ba
\frac{\omega}{2} \mbox{\boldmath$A$} (\xi) = 
\left( \Lambda_\gamma \F_\gamma \right)_{\alpha\beta} A_\beta (\xi) \, T_\alpha \, ,
\label{eigenequation} 
\ea
where the eight-dimensional matrices \( (\F_\alpha)_{\beta\gamma} \equiv - i f_{\alpha\beta\gamma} \) represent the adjoint representation of the $SU(3)$ Lie-algebra.

It is straightforward but very tedious to determine all the eigenvalues and the corresponding eigenvectors of the above eigenvalue equation (\ref{eigenequation}).
We shall only report the final results, but skip all the detailed derivations. 
It is found that the secular determinate of the matrix equation (\ref{eigenequation}) gives eight real eigenvalues:
\ba
\omega^2 
\left( 
\omega^6 - 6 B_2 \, \omega^4 + 9 B_2^2 \, \omega^2 - 4 (B_2^3 - 3 B_3^2)  
\right) = 0,
\la{secularequation}
\ea 
where \( B_2 = \Lambda_\alpha \Lambda_\alpha \) and \( B_3 = d_{\alpha\beta\gamma} \, \Lambda_\alpha \Lambda_\beta \Lambda_\gamma \) are identified with quadratic and cubic Casimir invariants constructed out of the Lie-algebra valued constant $\Lambda_\alpha$.
Note that two of the eigenvalues vanish in (\ref{secularequation}), whereas the other six are nonzero in general. 
We denote the six nonzero eigenvalues by \( \pm \, \omega_1, \pm \, \omega_2, \pm \, \omega_3 \) and choose them to be
\ba
\omega_1 &=& 2 \sqrt{B_2} \, \cos \frac{\phi}{6} ,
\nonumber \\
\omega_2 &=& - 2 \sqrt{B_2} \, \sin \frac{1}{6} (\pi + \phi) ,
\la{omega123} \\
\omega_3 &=& - 2 \sqrt{B_2} \, \sin \frac{1}{6} (\pi - \phi) ,
\nonumber
\ea
where \( \phi \equiv \cos^{-1} \left( 1 - \frac{6 B_3^2}{B_2^3} \right) \).
The byproduct of such an arrangement (\ref{omega123}) is that the eigenvalues 
$\omega_1$, $\omega_2$, and $\omega_3$ satisfy the following neat relations  
\ba
\omega_1 + \omega_2 + \omega_3 = 0,
\la{omegasummation} \\  
\omega_i \, \omega_j = \omega_k^2 - 3 B_2 .
\la{omegaproduct}
\ea
Here, the subscripts $i$, $j$, and $k$ in the latter equation (\ref{omegaproduct}) all are different and take values in (123).

In order to construct all of the eight eigenvectors of the eigenvalue equation (\ref{eigenequation}), we further define seven other Lie-algebra valued constants $\mbox{\boldmath$\Omega$}$, $\mbox{\boldmath$\Lambda_d$}$, $\mbox{\boldmath$\Omega_d$}$, $\mbox{\boldmath$\Lambda_f$}$, $\mbox{\boldmath$\Omega_f$}$, $\mbox{\boldmath$\Phi$}$, and $\mbox{\boldmath$\Delta$}$ that can be established from the known Lie-algebra constant $\mbox{\boldmath$\Lambda$}$. 
Among them, the first five Lie-algebra constants are given by
\ba
\mbox{\boldmath$\Omega$} 
&=& \Omega_\alpha T_\alpha 
= d_{\alpha\beta\gamma} \, \Lambda_\beta \, \Lambda_\gamma \, T_\alpha ,
\nonumber \\
\mbox{\boldmath$\Lambda_d$} 
&=& d_{8\alpha\beta} \, \Lambda_\beta \, T_\alpha ,
\nonumber \\
\mbox{\boldmath$\Omega_d$} 
&=& d_{8\alpha\beta} \, \Omega_\beta \, T_\alpha ,
\la{basisconstant1} \\
\mbox{\boldmath$\Lambda_f$} 
&=& i \, f_{8\alpha\beta} \, \Lambda_\beta \, T_\alpha ,
\nonumber \\
\mbox{\boldmath$\Omega_f$} 
&=& i \, f_{8\alpha\beta} \, \Omega_\beta \, T_\alpha .
\nonumber
\ea
In addition, the sixth constant is given by
\ba
\mbox{\boldmath$\Phi$} = i \, f_{i j \alpha} \, \Lambda_i \, \Omega_j \, T_\alpha ,
\la{basisconstant2}
\ea
where $i,j = 1,2,3$ belong to the $SU(2)$ subalgebra of $SU(3)$. 
The last one, with $\delta_{\alpha\beta}$ being the eight-dimensional kronecker delta, is
\ba
\mbox{\boldmath$\Delta$} = \delta_{8 \alpha} \, T_\alpha.
\la{basisconstant3}
\ea
With the definition of these Lie-algebra valued constants (\ref{basisconstant1}), (\ref{basisconstant2}), and (\ref{basisconstant3}), the eigenvectors of (\ref{eigenequation}) can be established in the following ways.
Two of the eigenvectors with vanishing eigenvalue are, respectively, shown to be
\ba
\mbox{\boldmath$\Lambda$} ~~~{\rm and}~~~
\mbox{\boldmath$\Sigma$} 
\equiv B_3 \mbox{\boldmath$\Lambda$} - B_2 \mbox{\boldmath$\Omega$} .
\la{eigenvector1}
\ea
As for the eigenvectors with nonvanishing eigenvalues, we designate $\mbox{\boldmath$V^i$}$ (for $i=1,2,3$) to be the corresponding eigenvector of the eigenvalue $\omega_i$. 
It is given by
\ba
\mbox{\boldmath$V^i$}
&=&
4 \left[ \Lambda_8 ( \omega_i^2 - B_2 )^2 - 3 B_3 \, \Omega_8 \right] \mbox{\boldmath$\Lambda$} -
12 \left[ \Omega_8 ( \omega_i^2 - B_2 ) + B_3 \, \Lambda_8 \right] 
\mbox{\boldmath$\Omega$} 
\nonumber \\
&& +
12 B_3 \, \omega_i^2 \, 
\mbox{\boldmath$\Lambda_d$} - 
6 \, \omega_i^2 ( \omega_i^2 - B_2 ) \,
\mbox{\boldmath$\Omega_d$}
\nonumber \\
&& -  
2 \, \omega_i ( \omega_i^2 - B_2 )
( \omega_i^2 - 2 B_2 - 2 \sqrt{3} \, \Omega_8 ) \,
\mbox{\boldmath$\Lambda_f$} -
4 \sqrt{3} \, \omega_i \left[ \Lambda_8 ( \omega_i^2 - B_2 ) - \sqrt{3} \, B_3 \right] 
\mbox{\boldmath$\Omega_f$} 
\nonumber \\
&& -
6 \sqrt{3} \, \omega_i ( \omega_i^2 - B_2 ) \,
\mbox{\boldmath$\Phi$} -
2 \left[ B_2 ( \omega_i^2 - B_2 ) ( \omega_i^2 - 2 B_2 ) - 6 B_3^2 \right]  \mbox{\boldmath$\Delta$} .
\la{eigenvector2}
\ea
Similarly, the corresponding eigenvector with eigenvalue $-\omega_i$ is just the hermitian conjugate of the eigenvector $\mbox{\boldmath$V^i$}$ (\ref{eigenvector2}), and will be denoted by $\mbox{\boldmath$\bar{V}^i$}$ for $i=1,2,3$.

It is not difficult to check the orthogonality conditions among these eigenvectors $\mbox{\boldmath$\Lambda$}$, $\mbox{\boldmath$\Omega$}$, $\mbox{\boldmath$V^i$}$, and $\mbox{\boldmath$\bar{V}^i$}$ ($i=1,2,3$). 
We exhibit the following nontrivial results 
\ba
\mbox{Tr} ( \mbox{\boldmath$\Lambda$} \, \mbox{\boldmath$\Lambda$} ) 
&=& 
\frac{1}{2} \, \Lambda^2 = \frac{1}{2} \, B_2 \, ,
\la{innerLambda} \\
\mbox{Tr} ( \mbox{\boldmath$\Sigma$} \, \mbox{\boldmath$\Sigma$} ) 
&=& 
\frac{1}{2} \, \Sigma^2 = \frac{1}{6} \, B_2 ( B_2^3 - 3 B_3^2 ) \, ,
\la{innerSigma} \\
\mbox{Tr} ( \mbox{\boldmath$V^i$} \, \mbox{\boldmath$\bar{V}^j$} ) 
&=& 
\frac{1}{2} \, \delta_{ij} \, (i \bar{i}) \, ,
\la{innerVi} 
\ea  
where $(i \bar{i}) \equiv V^i_\alpha \bar{V}^i_\alpha$ is the inner product of the eigenvector $\mbox{\boldmath$V^i$}$. 
Its explicit expression is irrelevant to our discussion, thus will not be shown here.
Besides, the rest of all other traces are found to be identically zero, for instances, we find that
\( \mbox{Tr} ( \mbox{\boldmath$\Lambda$} \, \mbox{\boldmath$\Sigma$} ) = 0 \),
\( \mbox{Tr} ( \mbox{\boldmath$\Lambda$} \, \mbox{\boldmath$V^i$} ) = 0 \),
\( \mbox{Tr} ( \mbox{\boldmath$\Sigma$} \, \mbox{\boldmath$V^i$} ) = 0 \),
\( \mbox{Tr} ( \mbox{\boldmath$V^i$} \, \mbox{\boldmath$V^j$} ) = 0 \), and so on.

To obtain the general solution of the Lie-algebra valued field $\mbox{\boldmath$A$}(\xi,\eta)$ fulfilling both (\ref{MCidentityansatz}) and (\ref{Motionansatz}), it is profitable to calculate the commutation relations among all of the eigenvectors. 
For the purpose of later convenience, we list the relevant commutators below 
\ba
\left[ \mbox{\boldmath$\Lambda$}, \mbox{\boldmath$\Sigma$} \right] = 0, 
\la{commutator1}
\ea
\ba
\left[ \mbox{\boldmath$\Lambda$}, \mbox{\boldmath$V^i$} \right] = 
\frac{\omega_i}{2} \, \mbox{\boldmath$V^i$}, 
~~~~~~
\left[ \mbox{\boldmath$\Lambda$}, \mbox{\boldmath$\bar{V}^i$} \right] = -\frac{\omega_i}{2} \, \mbox{\boldmath$\bar{V}^i$}, 
\la{commutator2}
\ea
\ba
\left[ \mbox{\boldmath$\Sigma$}, \mbox{\boldmath$V^i$} \right] = 
\frac{\omega_i}{2} \, {\cal P}(i) \, \mbox{\boldmath$V^i$}, 
~~~~~~
\left[ \mbox{\boldmath$\Sigma$}, \mbox{\boldmath$\bar{V}^i$} \right] =
-\frac{\omega_i}{2} \, {\cal P}(i) \, \mbox{\boldmath$\bar{V}^i$}, 
\la{commutator3}
\ea 
\ba
\left[ \mbox{\boldmath$V^i$}, \mbox{\boldmath$\bar{V}^j$} \right] = 
\delta_{ij} \, \frac{\omega_i}{2} \, (i \bar{i}) 
\left(
\frac{1}{B_2} \, \mbox{\boldmath$\Lambda$} + 
\frac{1}{\Sigma^2} \, {\cal P}(i) \, \mbox{\boldmath$\Sigma$}  
\right),
\la{commutator4}
\ea 
\ba
\left[ \mbox{\boldmath$V^i$}, \mbox{\boldmath$V^j$} \right] = 
\epsilon_{ijk} \, \frac{f(123)}{(k \bar{k})} \, \mbox{\boldmath$\bar{V}^k$}, 
~~~~~~
\left[ \mbox{\boldmath$\bar{V}^i$}, \mbox{\boldmath$\bar{V}^j$} \right] = 
- \epsilon_{ijk} \, \frac{f(123)}{(k \bar{k})} \, \mbox{\boldmath$V^k$} ,
\la{commutator5}
\ea 
where \( f(123) \equiv \frac{1}{2}\sqrt{ (1 \bar{1}) (2 \bar{2}) (3 \bar{3}) } \).
The function ${\cal P}(i)$ appearing in equations (\ref{commutator3}) and (\ref{commutator4}) is defined as
\ba
{\cal P}(i) = \left( B_3 - 2 B_2 \frac{(\Lambda i \bar{i})}{(i \bar{i})} \right), 
\la{Pfunction}
\ea
with $(\Lambda i \bar{i})$ $\equiv$ 
$d_{\alpha\beta\gamma} \, \Lambda_\alpha V^i_\beta \bar{V}^i_\gamma$.
It is stressed that the following identities for the function ${\cal P}(i)$ (\ref{Pfunction}) can be proved upon the use of Jacobi identities of the eigenvectors (\ref{eigenvector1}) and (\ref{eigenvector2})
\ba
\sum_{i=1}^3 {\cal P}(i) = 3 \, B_3 ~~~{\rm and}~~~~
\sum_{i=1}^3 \omega_i  \, {\cal P}(i) = 0 . 
\la{Pidentities}
\ea

Let us give a brief summary on what we have demonstrated concerning the solutions of the simplified Maurer-Cartan identity (\ref{MCidentityansatz}). 
Since it is a linear equation in the Lie-algebra valued field $\mbox{\boldmath$A$} (\xi,\eta)$, we are lead to determine the eigenvalues and the corresponding eigenvectors in the equation (\ref{eigenequation}). 
To this problem, the eigenvalues are the roots of the secular equation (\ref{secularequation}) and the nonvanishing roots are $\pm \, \omega_i$ for $i=1,2,3$ as shown in (\ref{omega123}), while the eigenvectors are illustrated in equations (\ref{eigenvector1}) and (\ref{eigenvector2}). 
As a result, the general solution for the Lie-algebra valued field $\mbox{\boldmath$A$} (\xi,\eta)$ satisfying (\ref{MCidentityansatz}) can be expressed in the form
\ba
\mbox{\boldmath$A$} (\xi,\eta) =  
f_\Lambda (\xi) \, \mbox{\boldmath$\Lambda$} + 
f_\Sigma (\xi) \, \mbox{\boldmath$\Sigma$} +
\sum_{i=1}^3 
\left[ 
f_i (\xi) \, e^{- \frac{i}{2} \omega_i \eta} \, \mbox{\boldmath$V^i$}  +
f_i^* (\xi) \, e^{\frac{i}{2} \omega_i \eta} \, \mbox{\boldmath$\bar{V}^i$} 
\right] .
\la{Ageneralsolution}
\ea
In this expression, the functions $f_\Lambda (\xi)$ and $f_\Sigma (\xi)$ are chosen to be real due to the hermiticity of the field $\mbox{\boldmath$A$} (\xi,\eta)$ (\ref{Adefinition}). 
In the same vein, the functions $f_i (\xi)$ for \( i =1,2,3 \) will be generically complex functions and $f_i^* (\xi)$ be the complex conjugate function of $f_i (\xi)$.
It is worth emphasizing that all of these functions are not arbitrary, but will be determined next by the Lagrange's equation of motion (\ref{Motionansatz}). 

\subsection{The Equation of Motion}

In (\ref{Ageneralsolution}), the expression for the general solution of the Lie-algebra valued field $\mbox{\boldmath$A$}(\xi, \eta)$ leaves us with eight functions undetermined. 
These functions are $f_\Lambda (\xi)$, $f_\Sigma (\xi)$, $f_i (\xi)$ and $f_i^* (\xi)$ ($i=1,2,3$).
In this subsection, we find the constraint relations among them by directly substituting the general solution (\ref{Ageneralsolution}) into the Lagrange's equation of motion (\ref{Motionansatz}). 
Explicitly, the constraint equations are 
\ba
&& \omega_k \, \frac{d}{d \xi} f_k (\xi) 
\nonumber \\
&=&
i \left[ 
\left( 2 \sigma - \frac{\omega_k^2}{2} 
\left( f_\Lambda (\xi) + f_\Sigma (\xi) \, {\cal P}(k) \right) \right) f_k (\xi) -
\frac{f(123)}{(k \bar{k})}
\sum_{i,j} \omega_i \, \epsilon_{ijk} \, f_i^* (\xi) \, f_j^* (\xi) 
\right] ,
\la{fequation1}
\ea
and the complex conjugate equation of (\ref{fequation1}).
Note that we have used the commutation relations 
(\ref{commutator1},\ref{commutator2}-\ref{commutator5}) in the derivation of the constraint equation (\ref{fequation1}).
The equations get further simplified if we define the functions
\ba
F_k (\xi) &=& \frac{1}{2} \, \sqrt{(k \bar{k})} \, f_k (\xi) ,
\la{Ffunction} \\
M_k (\xi) &=& 2 \sigma -\frac{\omega_k^2}{2} 
\left( f_\Lambda (\xi) + f_\Sigma (\xi) \, {\cal P}(k) \right) .
\la{Mfunction}
\ea
Then, using (\ref{Ffunction}) and (\ref{Mfunction}) to obtain this equation 
\ba
\omega_k \, \frac{d}{d \xi} \, F_k (\xi) = i \, 
\left( 
M_k (\xi) \, F_k (\xi) - 
\sum_{i,j} \omega_i \, \epsilon_{ijk} \, F_i^* (\xi) F_j^* (\xi) 
\right) 
\la{fequation2}
\ea
and its complex conjugate equation, where \( k=1,2,3 \). 

The equation (\ref{fequation2}) and its complex conjugate equation are of great interest by themselves, because they represent the Euler's equations of an asymmetrical top rotating in a complex three-dimensional space.  
In other words, they describe the rotational motion of an $SO(3,\C)$ top.
The three principal moments of inertia of the top are given by $\omega_1$, $\omega_2$, and $\omega_3$ (\ref{omega123}) and presumably all different. 
Observe that there are negative moments of inertia so that the total energy of the $SO(3,\C)$ top may not be positive definite.  
The angular velocity (complex) vector is defined by 
\( \vec{F}(\xi) = (F_1(\xi), F_2(\xi), F_3(\xi)) \).
Moreover, the top is not of free rotation, but rotating under the influence of the external torque, {\it i.e.}, the term \( M_k (\xi) \, F_k (\xi) \) appearing in (\ref{fequation2}).
As we will demonstrate later, although the external torque is presented in the Euler's equations, the problem is still integrable theoretically. The conservation laws can be established from the integrals of the Euler's equations. The discussion on the generic solutions of (\ref{fequation2}) is reported in the following paragraphs.
 
First of all, let the angular velocity component $F_k (\xi)$ $(k=1,2,3)$ be of the form
\ba
F_k (\xi) = G_k (\xi) \,
\exp \left( i \int^\xi d \xi' \frac{M_k (\xi')}{\omega_k} \right), 
\la{Fdefinition}
\ea
and use this to reformulate the Euler's equation of the top (\ref{fequation2}) as
\ba
\omega_k \, \frac{d}{d \xi} \, G_k (\xi) = -i \, 
\sum_{i,j} \omega_i \, \epsilon_{ijk} \, G_i^* (\xi) \, G_j^* (\xi) \, e^{-i \alpha \, \xi}.
\la{Gequation}
\ea
The parameter $\alpha$ in (\ref{Gequation}) is a constant and defined by
\ba 
\alpha \equiv \sum_{i=1}^3 \frac{M_i (\xi)}{\omega_i} = 
- \frac{3 B_2}{\sqrt{B_2^3-3B_3^2}} \, \sigma ,
\la{alphadefinition}
\ea 
where we have used (\ref{omegasummation}), (\ref{omegaproduct}), and the second identity of (\ref{Pidentities}).

Since $G_k (\xi)$, for $k=1,2,3$, are generally complex functions, we rewrite them using the polar coordinate representation
\ba
G_k (\xi) \equiv g_k (\xi) \, e^{i \, \theta_k (\xi)},
\la{Gpolar}
\ea
where by construction the modular function $g_k (\xi)$ and the angular function $\theta_k (\xi)$ are both real.
Then the application of (\ref{Gpolar}) to the top equation (\ref{Gequation}) and its complex conjugate equation results in differential equations fulfilled separately by $g_k (\xi)$ and $\theta_k (\xi)$ as
\ba
\frac{d}{d \xi} \, g_k^2 (\xi) 
&=& 
\frac{2}{\alpha}  \left(\frac{\omega_i - \omega_j}{\omega_k} \right) 
\frac{d}{d \xi} \, \Theta (\xi) ,
\la{gequation} \\
\frac{d}{d \xi} \, \theta_k (\xi) 
&=& 
- \left( \frac{\omega_i - \omega_j}{\omega_k} \right) 
\frac{\Theta (\xi)}{g_k^2 (\xi)} .
\la{thetaequation}
\ea
In the above equations, the subscripts $i$, $j$, and $k$ take values in the cyclic permutation set of (123). 
The explicit expression of the function $\Theta (\xi)$ appearing in (\ref{gequation}) and (\ref{thetaequation}) takes the form
\ba
\Theta (\xi) = \frac{1}{2}
\left[
G_1 (\xi) \, G_2 (\xi) \, G_3 (\xi) \, e^{i \alpha \, \xi} + 
G_1^* (\xi) \, G_2^* (\xi) \, G_3^* (\xi) \, e^{-i \alpha \,\xi}
\right] .
\la{Thetafunction}  
\ea

It is apparent that from (\ref{gequation}) we obtain three conserved quantities, the constants of motion. 
Let us them denoted by $C_k$ ($k=1,2,3$). 
They are known as the integrals of Euler's equations
\ba
g_k^2 (\xi) - \frac{2}{\alpha} 
\left( \frac{\omega_i - \omega_j}{\omega_k} \right) \Theta (\xi) = C_k.
\la{modularsolution}
\ea
So, the modular functions $g_k (\xi)$ (\ref{Gpolar}) are not independent at all, but can be expressed in terms of the function $\Theta (\xi)$ (\ref{Thetafunction}).
In the same vein, the angular functions $\theta_i (\xi)$ defined in (\ref{Gpolar}) are all correlated through the same $\Theta (\xi)$ function and can be represented as the integration:
\ba
\theta_k (\xi) = - 
\int^\xi d \xi' 
\left[ 
\frac{2}{\alpha} + \frac{\omega_k}{\omega_i - \omega_j} \frac{C_k}{\Theta (\xi')}
\right]^{-1}. 
\la{angularsolution}
\ea

As a result, all the modular and angular functions $g_k (\xi)$ and $\theta_i (\xi)$ are expressible in terms of the single function $\Theta (\xi)$. 
However, we are still left with this function yet to be determined.
To determine the differential equation that the $\Theta (\xi)$ function satisfies, we take the differentiation of the equation (\ref{Thetafunction}). 
It is found, with the use of the polar representation (\ref{Gpolar}), that 
\ba
&& \frac{d}{d \xi} \Theta (\xi) = - {\alpha} \times
\nonumber \\
&& ~~ 
\Bigg\{ 
\left[ \frac{2}{\alpha} \frac{\omega_2 - \omega_3}{\omega_1} \, \Theta + C_1 \right]
\left[ \frac{2}{\alpha} \frac{\omega_3 - \omega_1}{\omega_2} \, \Theta + C_2 \right] 
\left[ \frac{2}{\alpha} \frac{\omega_1 - \omega_2}{\omega_3} \, \Theta + C_3 \right] -
\Theta^2
\Bigg\}^{1/2} .
\la{Thetadifferential}
\ea
The integration of this equation generally renders the $\Theta (\xi)$ as a generic function of the variable $\xi$, though the precise form of this function is unknown and troublesome.   
Anyhow, the problem (\ref{fequation2}) of the spinning $SO(3,\C)$ top in the presence of external torque is exactly solvable, as we have claimed before.
Note that the integral (\ref{Thetadifferential}) will be reduced to its simplest expression if the conserved quantities $C_i$ ($i=1,2,3$) (\ref{modularsolution}) are chosen to satisfy
\ba
\sum_{i,j,k}
\left[ 
C_i \, \frac{(\omega_k - \omega_i)(\omega_i - \omega_j)}{\omega_j \, \omega_k} 
\right]
= \frac{\alpha^2}{4} ,
\la{simplestcase}
\ea
where $i$, $j$, and $k$ take values in the cyclic permutations of (123).
Under such a choice, the quadratic term in $\Theta (\xi)$ vanishes inside the radical sign on the right-hand-side of equation (\ref{Thetadifferential}). 
For this time, the integration of (\ref{Thetadifferential}) will give the function $\Theta (\xi)$ as a Weierstrass elliptic function, which is more familiar to us.
Since the complete analysis of the solutions for the $\Theta (\xi)$ function is out of the scope of this paper, we shall not pursue it further here. 

We thereby give a brief note on the general solution of the Lie-algebra valued field $\mbox{\boldmath$A$}(\xi,\eta)$ satisfying both equations (\ref{MCidentityansatz}) and (\ref{Motionansatz}). 
It was mentioned in the previous subsection that the solution of $\mbox{\boldmath$A$}(\xi,\eta)$ in (\ref{Ageneralsolution}) has eight functions needed to be determined by the Lagrange's field equation (\ref{Motionansatz}). 
Among them, the functions $f_i (\xi)$ and $f_i^* (\xi)$, related to the complex angular velocity $F_i (\xi)$ (\ref{fequation2}) of the rotating $SO(3,\C)$ top, can be solved in terms of the $\Theta (\xi)$ function (\ref{Thetadifferential}). 
However, both functions $f_\Lambda (\xi)$ and $f_\Sigma (\xi)$ remain to be arbitrary since the equation of motion yields no consequence on them at all.
At last, the final result of the field $\mbox{\boldmath$A$}(\xi,\eta)$ is expressed in this form
\ba
&&
\mbox{\boldmath$A$} (\xi,\eta) =  
f_\Lambda (\xi) \, \mbox{\boldmath$\Lambda$} + 
f_\Sigma (\xi) \, \mbox{\boldmath$\Sigma$} +
\nonumber \\
&&
\sum_{i=1}^3 
\left[ 
\frac{2 g_i (\xi)}{\sqrt{(i \bar{i})}} \, 
\exp \left[ - \frac{i}{2} \left( \omega_i \, \eta + \alpha \, \xi \right) +
i \int^\xi d \xi'
\left( \frac{M_i (\xi')}{\omega_i} + \frac{\alpha}{2} \frac{C_i}{g_i^2 (\xi')} \right) 
\right]
\mbox{\boldmath$V^i$} + ({\rm h.c.})
\right]
\la{Ageneralsolution1}
\ea
where (h.c.) denotes the hermitian conjugation.
$M_i (\xi)$ is related to the functions $f_\Lambda (\xi)$ and $f_\Sigma (\xi)$ in (\ref{Mfunction}), while the function $g_i (\xi)$ is related to the $\Theta (\xi)$ function in (\ref{modularsolution}).

\section{\la{DISCUSSIONS AND CONCLUSIONS}
             DISCUSSIONS AND CONCLUSIONS}

The Lie-algebra valued field $\mbox{\boldmath$A$}(\xi,\eta)$ given in (\ref{Ageneralsolution1}) contains two arbitrary functions $f_\Lambda (\xi)$ and $f_\Sigma (\xi)$. 
Thus, the general expression of $\mbox{\boldmath$A$}(\xi,\eta)$ that solves both equations (\ref{MCidentityansatz}) and (\ref{Motionansatz}) specifies a class of exact solutions of the $SU(3)$ Skyrme model.
These solutions are not topological solitons, since the baryon number current defined in (\ref{baryoncurrent}) and the baryon number in (\ref{baryonnumber}) both are vanishing under the ansatz (\ref{HYansatz}) or equivalently (\ref{leftcurrent1}).
In addition, the solutions (\ref{Ageneralsolution1}) that we obtain are neither the pure superposition of plane waves in contrast to the $SU(2)$ Skyrme model \cite{Hirayama}, but in the simplest situation are the complicated combination of plane waves and the Weierstrass elliptic functions.  

It is stressed that the characteristics of the Lie-algebra valued field $\mbox{\boldmath$A$}(\xi,\eta)$ shown in (\ref{Ageneralsolution1}) is somewhat quite general and can be applied to other higher rank groups, such as the group $SU(N)$ for \( N \ge 4 \).
Take the $SU(N)$ group as an example, there would be $N-1$ arbitrary functions, those like $f_\Lambda (\xi)$ and $f_\Sigma (\xi)$, in the final expression of the field $\mbox{\boldmath$A$}(\xi,\eta)$.
Also, there would exist generalized Euler's equations, generalizing the equation (\ref{fequation2}), which generically describe an asymmetric top rotating in the presence of generalized external torque in the complex \( \frac{1}{2} N(N-1) \)-dimensional space.  
However, the determination of the eigenvalues and eigenvectors of the Maurer-Cartan identity (\ref{MCidentityansatz}) for the $SU(N)$ group remains a difficult task. 

The class of solutions that are established here for the $SU(3)$ Skyrme model can be easily reduced to the case of the $SU(2)$ model \cite{Hirayama}. 
For this purpose, let us examine the both equations (\ref{Ageneralsolution}) and (\ref{fequation1}) under the $SU(2)$ reduction. 
Since the adjoint representation of the $SU(2)$ algebra is three-dimensional, we first set \( f_\Sigma (\xi) = 0 \) and \( f_i (\xi) = 0 \) (for $i=2,3$) in the equation (\ref{Ageneralsolution}). 
Then, for the equation (\ref{fequation1}), the only surviving function $f_1 (\xi)$ under the $SU(2)$ reduction fulfills the simple differential equation 
\ba
\omega_1 \, \frac{d}{d \xi} f_1 (\xi) =
i \left( 2 \sigma - \frac{\omega_1^2}{2} \, f_\Lambda (\xi) \right) f_1 (\xi) .
\la{fequationSU2}
\ea
After the result of integration of this equation (\ref{fequationSU2}) is substituted into the equation (\ref{Ageneralsolution}), we recover the general solution of the $SU(2)$ Lie-algebra valued field $\mbox{\boldmath$A$}(\xi,\eta)$ as
\ba
\mbox{\boldmath$A$} (\xi,\eta) =  
f_\Lambda (\xi) \, \mbox{\boldmath$\Lambda$} + 
\left\{
\exp
\left[ i \, \frac{2 \sigma}{\omega_1} \, \xi
- i \, \frac{\omega_1}{2} 
\left( \eta + \int^\xi d \xi' \, f_\Lambda (\xi') \right) 
\right]
\mbox{\boldmath$V^1$}  +
({\rm h.c.})
\right\} .
\la{AgeneralsolutionSU2}
\ea
It is noted that this is the precise expression reported in the paper \cite{Hirayama} up to normalization constants. 

We conclude the paper by discussing an extreme case on the solution of the field $\mbox{\boldmath$A$}(\xi,\eta)$ (\ref{Ageneralsolution1}), where the two arbitrary functions $f_\Lambda (\xi)$ and $f_\Sigma (\xi)$ are set to zero for simplicity. 
Moreover, we take the parameter $\sigma = 0$, that is, the pion decay constant \( F_\pi = 0 \). 
Such a choice is called extreme because we are analyzing a very particular set of solutions, in which the Skyrme term is the only term presented in the Skyrme model (\ref{skyrme2}). 
For this extreme case, the general solutions of the Lie-algebra valued field $\mbox{\boldmath$A$}(\xi,\eta)$ are obtained by the direct replacement of $\sigma$, $f_\Lambda (\xi)$ and $f_\Sigma (\xi)$ in equation (\ref{Ageneralsolution1}). 
However, let us focus on the equation (\ref{fequation2}) instead.
It is reduced to this equation
\ba
\omega_k \, \frac{d}{d \xi} \, F_k (\xi) = - i \, 
\sum_{i,j} \omega_i \, \epsilon_{ijk} \, F_i^* (\xi) F_j^* (\xi) .
\la{fequationextreme}
\ea
Obviously, the equation (\ref{fequationextreme}) and its complex conjugate equation describe an asymmetrical top that is freely rotating in the complex three-dimensional space.  
Both equations together represent the Euler's equations of the rotational $SO(3,\C)$ top.
The complex angular velocity vector is written as before by
\( \vec{F}(\xi) = (F_1(\xi), F_2(\xi), F_3(\xi)) \).
If we denote the complex angular momentum by $\vec{L} (\xi)$ = 
\( \left( L_1 (\xi), L_2 (\xi), L_3 (\xi) \right) \) $\equiv$ 
\( \left( \omega_1 F_1, \omega_2 F_2, \omega_3 F_3 \right) \), then two integrals of the Euler's equations (\ref{fequationextreme}) are known already from the laws of conservation of energy and angular momentum:   
\ba
\frac{|L_1|^2}{\omega_1} + \frac{|L_2|^2}{\omega_2} + \frac{|L_3|^2}{\omega_3} &=& 2 E ,
\la{energyconservation} \\
|L_1|^2 + |L_2|^2 + |L_3|^2 &=& L^2 ,
\la{momentumconservation}
\ea
where the energy $E$ and the magnitude $L$ of the angular momentum are given constants.
The third conserved quantity for this freely spinning top can be chosen, for example, as the function 
\ba
\Theta = \frac{1}{2}
\left[
F_1 (\xi) \, F_2 (\xi) \, F_3 (\xi) + F_1^* (\xi) \, F_2^* (\xi) \, F_3^* (\xi)
\right] .
\la{Thetafunctionextreme}  
\ea
The $\Theta$ function in (\ref{Thetafunctionextreme}) is a constant of motion since \( \frac{d}{d \xi} \Theta = 0 \) in the extreme situation.
Consequently, the problem (\ref{fequationextreme}) of the freely rotating top is again exactly solvable.
The final form of the solutions is expressed in terms of an integral formulation, which is similar to that appearing in the equation (\ref{Thetadifferential}).    
By the same token, this integral is reduced to another Weierstrass elliptic function in the simplest circumstance. 

\vskip 1cm

The author is grateful to C. R. Lee and M. H. Tu for useful discussions.
This work was supported in part by Taiwan's National Science Council Grant No. 91-2112-M-194-009.

\end{document}